\documentclass[aps,prapplied,twocolumn,floatfix]{revtex4-2}
\usepackage[arrowdel]{physics}
\usepackage{graphicx,psfrag}
\usepackage[dvipsnames]{xcolor}
\usepackage{amssymb}
\usepackage{mathtools} 
\usepackage{textgreek} 
\usepackage{comment}

\newcommand{\pref}[1]{(\ref{#1})}
\newcommand{\epref}[1]{Eq.~(\ref{#1})}
\newcommand{\eprefs}[2]{Eqs.~(\ref{#1},\ref{#2})}
\newcommand{\half}{\frac 12}
\newcommand{\erfc}[1]{\text{erfc}\left[ #1 \right]}
\newcommand{\ie}{\textit{i.e. }}

\newcommand{\dzitr}{\dot z(-\infty) T/R}

\newcommand{\phio}{\phi_o}
\newcommand{\mcL}{{\mathcal L}}

\newcommand{\Thetarho}{\Theta_\rho}

\begin{document}
\title{Current-induced mechanical torque in chiral molecular rotors}
\author{Richard Koryt\'ar}
\affiliation{Department of Condensed Matter Physics, Faculty of Mathematics and Physics, Charles University, Ke Karlovu 5, 12116 Praha 2, Czech Republic}
\author{Ferdinand Evers}
\affiliation{Institute of Theoretical Physics, University of Regensburg, D-93050 Regensburg, Germany}

\date{\today}

\begin{abstract}
A great endeavor has been undertaken to engineer molecular rotors operated by
an electrical current. A frequently met operation principle is the transfer of
angular momentum taken from the incident flux. In this paper we present an
alternative driving agent that works also in situations where angular momentum
of the incoming flux is conserved. This situation arises typically with
molecular rotors that exhibit an easy axis of rotation. For quantitative
analysis we investigate here a classical model, where molecule and wires are
represented by a rigid curved path. We demonstrate that in the
presence of chirality the rotor generically undergoes a directed motion,
provided that the incident current exceeds a threshold value. Above threshold,
the corresponding rotation frequency (per incoming particle current) for
helical geometries turns out to be $2\pi m/M_1$, where $m/M_1$ is the ratio of
the mass of an incident charge carrier and the mass of the helix per winding
number. 
\end{abstract}

\maketitle
\section{Introduction}
Experiments employing the scanning-tunneling microscopy 
(STM) have achieved a directed rotation of a molecule controlled by an electrical current.  
Correspondingly, realizations of molecular switches and rotors have been reported, 
\cite{Lensen2012,Perera2013,Schaffert2013,Ohmann2015, Simpson2019,
JasperTonnies2020,Wu2020,Eisenhut2021,Ren2020}, with potential relevance for future molecular technologies. 
%

The theory describing the working principle of such {\em molecular motors} 
often employs angular Langevin equations \cite{Stolz2020,Ren2020}. The method has been established by H\"anggi
\cite{Hangi2009}, Astumian \cite{Astumian2016} and collaborators in the context of
Brownian motors. It describes the dynamics of a classical angular variable $\vartheta$
that is subject to a "ratchet"-type of potential in the presence of a (phenomenologically treated) driving torque.
\emph{Ab-initio} expressions for the current induced torques have been obtained within the non-equilibrium Green's function formalism  
\cite{Bode2011, Todorov2014}. The current excites a variety of molecular vibrational modes, rendering the atomistic  analysis of the torque very complex (see Ref. \cite{JasperTonnies2020} for an \emph{ab-initio} calculation of the vibrations). 

\begin{figure}[b]
    \centering
    \includegraphics[width=.3\columnwidth]{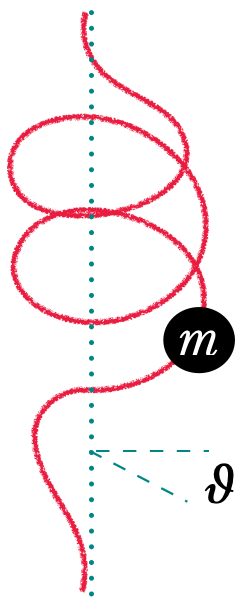}%
    \hfill %
    \includegraphics[width=.6\columnwidth]{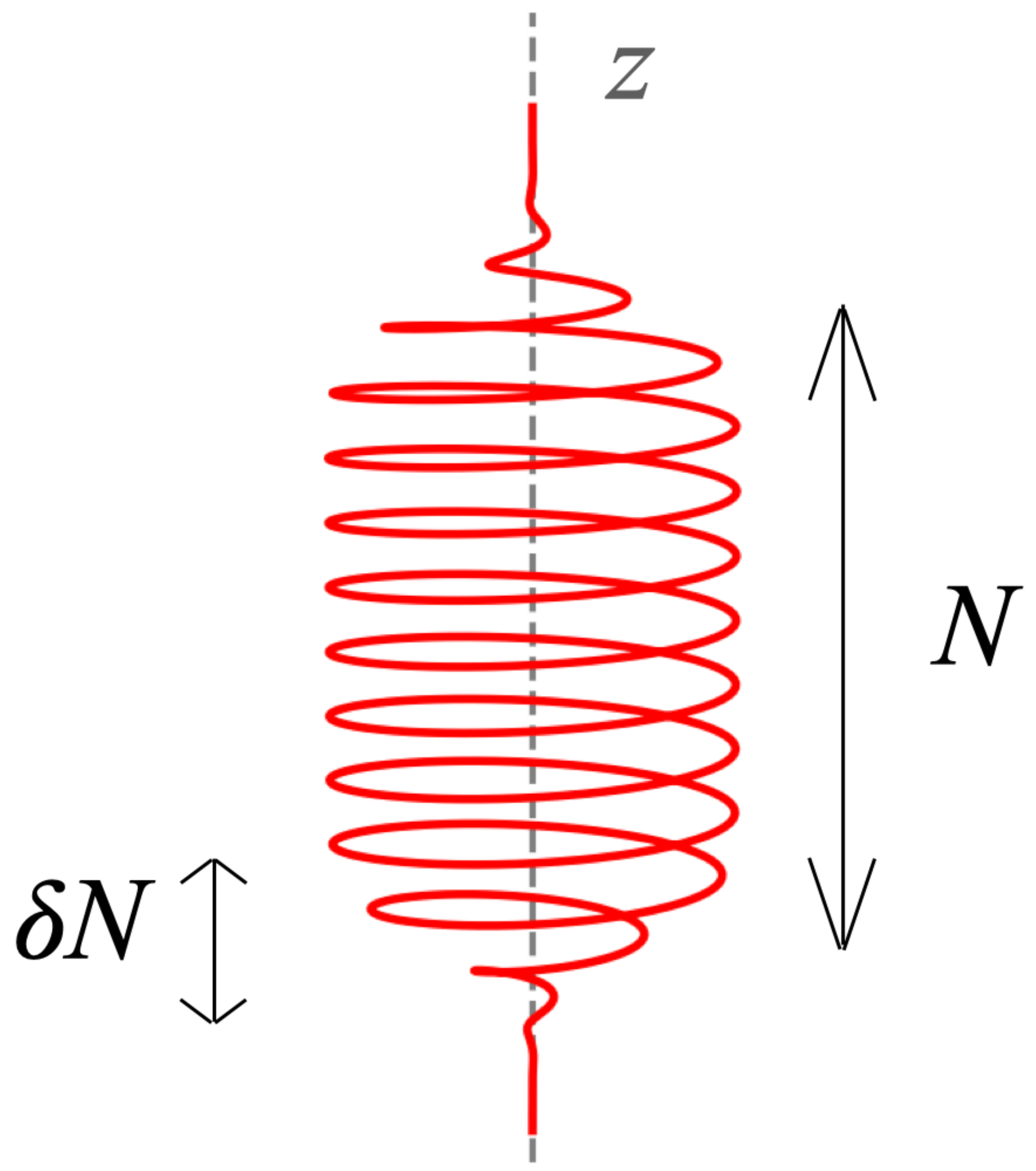}
    \caption{Left: A particle of mass $m$ is constrained to move along a path (red curve). 
    An axis is assumed, so that the initial and final radius of the path with respect to the axis is zero.
    Furthermore, the path is allowed to rotate (angle $\vartheta$) around the axis.
    Right: $N$-helix which smoothly evolves from a straight line to its radius $R$, parameterized by the formulas in the Appendix.
    Here, the number of turns is $N=10$ and the full radius sets in from zero after $\delta N =1$ turns.
    }
    \label{fig:scheme}
\end{figure}

To bring about a controlled unidirectional rotation in the STM setup requires a degree of symmetry breaking. There are two typical situations: either the molecule by itself exhibits a handedness (chirality) or chirality is imposed by the geometry of the molecular junction, for examples see \cite{JasperTonnies2020} or  \cite{Stolz2020}. 
The purpose of this article is to provide a qualitative description of the current-induced mechanical torque within a toy model framework.  

We consider a classical model of the molecular rotor, where the molecule is modeled as a one-dimensional curve ("molecular wire") that guides the flow of the charge carriers. 
 (see Fig.~\ref{fig:scheme} (left) for illustration). 
The motion of the particle along the molecule obeys 
Lagrangian dynamics. The wire can rotate around a given axis with angle $\theta$. The torque driving the rotation is provide by the backaction of moving particle. In the absence of a potential $V(\theta)$ angular momentum is conserved. 

The main outcome of this work is that the wire rotates even if the the net transfer of angular momentum of the transmitted particles is zero. 
The operation principle is that the particle exerts a torque when entering and leaving the molecular wire. Even if both exactly compensate, the wire rotates while the particle travels along, so that each transmitted particle results in a shift $\delta \theta$. 
 This operational principle is different than an earlier reported \cite{Kral2005} where an electric field was needed to continuously accelerate the electrons while they travel along a helical wire. 
When the rotation of the molecule is hindered by a potential barrier $V(\theta)$, we find that the mass current needs to overcome a threshold for the wire to rotate.  The resulting time-averaged angular velocity is time independent and directional for all supercritical currents.
%

Finally, we consider a situation where the net torque exerted by the transmitted particle does not vanish.. 
In this situation the rotation trivially appears due to the momentum transfer ("garden-hose effect").
This situation represents molecular junctions where the incoming or outgoing current can carry a non-vanishing angular momentum.
We present the characteristics of the crossover between both regimes.

Summarizing, our work provides  insights into the operation principles of molecular rotors, specifically the velocity-current characteristics and threshold currents. Our results can support the design of nanoscale mechanical devices.

\section{Model \label{sec:model}}
\subsection{Model geometry (kinematics)}
Our classical model contains a particle (mass $m$) moving on a rigid path, which can
rotate around an axis, see left part of Fig.~\ref{fig:scheme}. The rotation angle of the path is denoted by $\vartheta$.
In absence of the rotation degree of freedom of the path,
the particle would experience constrained dynamics. With the rotation allowed, the motion
of the particle can exert a torque on the path. Conversely, the dynamics of the path around
its angle affects the passage of the particle.

The trajectory (path) at rest ($\dot \vartheta = 0$, $\vartheta = 0$) will
be expressed parametrically in a cylindrical coordinate system:
\begin{align}
\label{eq:path}
(\rho(s), \phio(s), z(s)). 
\end{align} 
The parameter $s$ could be the distance along the path; for the purpose of this work it is not
required.
For simplicity, we further stipulate that $z(s)$ is monotonously increasing with $s$,
and that the trajectory never intersects itself.

The model contains two dynamical variables, the degree of freedom of the particle, $s(t)$, 
and $\vartheta(t)$, the latter being the
angle of the path with respect to a static coordinate system.
Our aim is to investigate the dynamics of $\vartheta$ under the condition that
the incoming and outgoing particles don't carry any angular momentum. We achieve this
by conditioning the path to have vanishing radius at its start and at the end,
\begin{equation}
\label{eq:zerorho}
    \rho(-\infty) = \rho(+\infty) = 0.
\end{equation}
Later on we also employ a path with a finite final radius, allowing for an angular momentum transfer. As we shall demonstrate, paths satisfying \epref{eq:zerorho} will still turn when subjected to particles, if the
path is chiral (lacks reflection symmetry). We shall employ a helical path,
with radius smoothly raising from zero, effecting $N$ turns and sinking at the end,
see Figure~\ref{fig:scheme}(b). The mathematical expression of the path can be found in the 
Appendix.
 
\subsection{Lagrangian dynamics}

We construct the equations of motion from a Lagrangian, $\mcL(\vartheta, \dot\vartheta, s,\dot s)$.

\subsubsection{Static path}
For a fixed path, $\vartheta = 0$, the Lagrangian of this model reduces to
the Lagrangian of a particle subject to a constraint,
\begin{align}
    \mcL_0(s,\dot s) &= \frac{m}{2} \left(\dot\rho^2 + (\rho\dot\phio)^2 + \dot z^2\right) \\
    &= \frac{m}{2} \left(\frac{\partial\rho}{\partial s}^2 
     + (\rho\frac{\partial \phio}{\partial s})^2 + \frac{\partial z}{\partial s}^2 
    \right)\dot s^2 .
    \label{lagrange1} 
\end{align}
Formally, as a consequence of the constraint the model adopts the form of a
free particle with an $s$-dependent mass.  Recalling the conservation of
energy, the formal integration of the Lagrangian \eqref{lagrange1} is trivial. 

\subsubsection{Dynamic path.}
To allow a dynamical rotational degree of freedom $\vartheta$ for the path, we 
now introduce the actual angle $\phi$ of the particle in a static cylindrical system, defined by
\begin{align}
    \phi = \phio + \vartheta
\end{align}
and introduce it in \epref{lagrange1}.

Without a particle on a path, the dynamics of the rotor will be governed by the kinetic energy
$\half \Theta \dot \vartheta^2$ and the potential energy $V(\vartheta)$.

The full Lagrangian of the coupled system becomes
\begin{align}
    \mcL(\vartheta, \dot\vartheta, s,\dot s) &= \frac{1}{2}\Theta \dot \vartheta^2  - V(\vartheta) +\\
    &{+} \frac{m}{2} \left(\frac{\partial\rho}{\partial s}^2\dot s^2 +
     \rho^2(\frac{\partial \phio}{\partial s}\dot s +\dot\vartheta)^2 + \frac{\partial z}{\partial s}^2 \dot s^2 
    \right)
    \label{eq:lagrangian} 
\end{align}

It provides two equations of motion (EOM), which are listed in the Appendix. We integrate the EOM
using a Runge-Kutta method, see Appendix.

\subsubsection{Basic parameters and scales}
The parameters that enter the coupled dynamical problem governed by the Lagrangian \pref{eq:lagrangian} are
\begin{itemize}
\item particle mass $m$
\item the definition of the path, \epref{eq:path}. It will be assumed,
      that the path has a characteristic radius $R \sim \text{max}\,\rho(s)$, which will serve as 
      a length scale. Helical paths are primarily distinguished by the number
      of turns $N$, which controls the time particle spends on the helix.
\item inertia tensor of the path $\Theta$. The latter can be expressed
      through the characteristic radius as $M R^2$, defining mass $M$.
      The ratio $\mu = m/M$ enters in the collision characteristics.
\item potential $V(\vartheta)$ that hinders the motion of the path
      (setting a preferred direction). The difference between the min and max
      is denoted by $\Delta V$. This is the energy scale that needs
      to be overcome when inducing an unbound rotation -- else one
      is trapped in the potential valley.
\item The above parameters of the path combine to give a time scale
      $T := \sqrt{\Theta / \Delta V}$, which equals $0.334$ times the
      period of small harmonic oscillations of the path without the particle,
      around the potential minimum. We shall use $T$ as a unit of time in our numerical results.
\item The initial velocity of the particle, at $s=-\infty$,
       denoted by $\dot z(-\infty)$, ''the impact velocity''. A suitable unit for the latter is $R/T$ and it is inversely proportional to the time spent in the curved path. 
\item The precise initial placement of the particle, $z(-\infty)$, is irrelevant,
       because the particle decouples from the path when $\rho = 0$.
\end{itemize}

\subsection{Conservation laws}
Consider a single collision event, with particle
starting at $s=-\infty$, passing through the rotor (where $\rho \ne 0$) and leaving
towards $s=\infty$\footnote{Projectiles with a very small energy can be,
in principle, reflected, but this regime is beyond the scope of this work.}.
\paragraph{Energy conservation.} 
If before the collision the path is at rest,
energy conservation implies that
\begin{equation}
\label{eq:energy}
 \Delta E = \left.\half \Theta\dot\vartheta^2 + V(\vartheta)\right|_{t=\infty} = \half m\left[\dot z(-\infty)^2 - \dot z(\infty)^2\right].
\end{equation}
This is a consequence of the invariance of the Lagrangian \pref{eq:lagrangian} with respect
to time translations. The right hand side of \epref{eq:energy} describes
the energy loss of the particle after the collision.  We shall focus on
the regime where the energy gain of the path, $\Delta E$, is small,
usually not higher than $\Delta V$. In the limit of fast impact
velocities, \epref{eq:energy} implies that the relative decrease of
the particle velocity after the collision is small.

\paragraph{Angular momentum conservation.}
When $V(\vartheta)=\text{const.}$, the Lagrangian \pref{eq:lagrangian} is invariant with respect to rotations. The total angular momentum
\begin{equation}
\label{eq:am}
    J = \Theta \dot \vartheta + m\rho^2 \dot\phi 
\end{equation}
is time independent. For paths satisfying \epref{eq:zerorho} $J$ equals the
angular momentum of the path before and after the collision.

\subsection{Path under a current: Stroboscopic dynamics}
We will also investigate a dynamics of the path when the particles come
sequentially, \textit{i.e.} the path under a current.
When the particles arrive to the path periodically, with period $\Delta t$,
the particle current reads
\begin{equation}
I(t) = \sum_{n=-\infty}^\infty \delta (t - n\Delta t).
\end{equation}
The time-averaged current reads $\langle I(t)\rangle = I = 1/\Delta t$.

We will assume for each incident particle identical initial conditions,
\textit{i.e} at each time $n\Delta t$ the same $z$ and $\dot z(-\infty)$.
However, the initial conditions for $\vartheta$ and $\dot\vartheta$ will be different,
corresponding to the dynamical state of the path. In between the sequential collisions, the path evolves under its independent
 equation of motion, $\Theta \ddot \vartheta = -V'(\vartheta)$.

\section{Results}
First, we demonstrate how a particle that does not carry any angular momentum can turn the path.
It is instructive to begin with the limit of full rotational invariance, when $V=0$, Sec.~\ref{sec:rotinv}, because conservation laws allow for
a straightforward integration of the EOM. Next, we treat analytically
the case $V\ne 0$ in the limit of fast projectiles in the so called
sudden approximation (SA) in Sec.~\ref{sec:sa}. We use the analytical
considerations as guiding principles for the analysis of the numerics
in Sec.~\ref{sec:numr1}.
After that we consider paths which allow for a finite angular momentum transfer in Sec.~\ref{sec:ghose}.

\subsection{\label{sec:rotinv}Rotational invariance}
When $V=0$, the dynamics of a single shot (collision) is entirely captured by angular momentum conservation.

Let us assume the path at rest before the collision. \epref{eq:am} with $J=0$ binds the change of the angle of the particle with the change of the angle of the path
(analogous with Keppler's law)
\begin{equation}
  -m\rho^2(s) \mathrm d\phi(s) = \Theta \mathrm d \vartheta.
\end{equation}
Integrating from $s=-\infty$ to $s=\infty$, we obtain
\begin{equation}
\label{eq:turn}
 -m 2A = \Theta \Delta \vartheta
\end{equation}
where on the left hand side $A = \half \int \rho^2\mathrm d\phi$ denotes the area described by the 'clock'
with a variable radius $\rho(s)$ during the passage of the particle.
For the $N$-helix, $A\approx \pi R^2 N$.
On the right hand side we obtained the change of the angle of the path.

Although the path can experience a turn, no angular velocity is generated after
the collision, as a consequence of \epref{eq:zerorho}.  The traversing particle
does exert a torque, however, when the torque is integrated over time, it
produces no net angular velocity.
The path  experiences a turn
in a preferred direction. For a general path,
the turn is finite, if the 'clock' area is finite. This situation can not be realized in paths with a spatial reflection plane or an inversion point located on the path. For paths with handedness (chirality), the sign of the turn is determined by the
 the sign of $A$, which, in turn, has the chirality sign.
 
Next, while still assuming $V=0$, we ponder three specific sectors centered around $\vartheta = 0, \pm 2\pi/3$, and investigate the conditions for a single particle to switch the $N$-helix from one sector to another.
The condition is that $\Delta\vartheta$ reaches $\pm 2\pi / 6$.
Combining with \epref{eq:turn}, we arrive at
\begin{equation}
\label{eq:sudden1}
N \frac mM = \frac 16
\end{equation}
(or more precisely, $A \frac mM = \frac 16 R^2 \pi$).
In the above formula, $M$ is proportional to $N$, so that the
required threshold particle mass $m$ is independent on the length
of the helix.

\subsection{\label{sec:sa}Broken rotational invariance: analytic considerations in the sudden
approximation}

We will be concerned with a situation in which the rotation of the path
will be hindered, in order to study switching.
We introduce a potential 
\begin{equation}
    \label{eq:v}
    V = \Delta V \sin^2(3\vartheta/2)
\end{equation}
 with three minima, separated by obstacles of the height $\Delta V$.
Three-state rotors have been reported recently in experimental Refs.~\cite{Stolz2020, JasperTonnies2020}.
Rotational invariance is broken and although there is energy conservation,
the equations of motion are difficult to treat analytically. However,
there is a limit where approximations are feasible.

\subsubsection{Single particle dynamics.}
If the passage time of the particle $\delta t$ is much smaller
 than the oscillation period $\approx T$, we may safely neglect the
 potential in the collision problem.
The formulas (\ref{eq:sudden1},\ref{eq:turn}) remain valid in this limit
and will serve us as a useful guide.
The above mentioned condition for the applicability of the formulas of the sudden approximation (SA)
can be formulated as
\begin{equation}
    \label{eq:condt}
T\ \gtrapprox\ \frac{L_N}{\dot z(-\infty)} \approx \frac{2\pi RN} {\dot z(-\infty)},
\end{equation}
where the passage time on the right side has been approximated from a uniform
motion of the projectile over the path length $L_N$ (where $\rho> 0$). For
an $N$-helix parameterized in the Appendix, 
\begin{equation}
    \label{eq:condt2}
  L_N  \approx 2\pi NR.  
\end{equation}

An important feature in the broken rotational invariance is that
the restoring torque gives the path an acceleration once the particle is gone.
The resulting motion can be, in general, bound to the potential minimum
or unbound, depending on the parameters. In the SA, the condition
that separates the two regimes is expressed by the \epref{eq:sudden1}.

\subsubsection{Helix under a current.}
A single particle may not cause a turn that is sufficient for an unbound
motion if the mass ratio $\mu = m/M$ is too low, for example. But the required critical turn can be effected if particles
arrive sequentially, \ie under a current $I$. According to the formula
\pref{eq:sudden1}, each particle induces an \textit{angular boost} as it passes.

If the current runs for a time $t$ (to be specified later), formula \pref{eq:sudden1} becomes
\begin{equation}
  \label{eq:switchi0}
   I t N \frac mM = \frac 16.
\end{equation}
It determines the minimum threshold mass current $I_m = mI$ required to perform a switch.
The last formula is applicable under specific conditions:
\begin{enumerate}
    \item \epref{eq:switchi}
    comes from the SA, demanding that the impact velocity is large enough, \epref{eq:condt}.
    \item The time between collisions should be much smaller than $T$ in order to silence the restoring torque:
     $I^{-1} \ll T$
    \item The switching time $t$ must also be much shorter than $T$, else the path likely
    performs an oscillatory motion against a displaced minimum.
\end{enumerate}
Our objective will be to determine the threshold mass current $I_m$. Therefore,
the condition \epref{eq:switchi0} from the SA can be written as
\begin{equation}
\label{eq:switchi}
    I_m T = \frac{M_1}6
\end{equation}
where $M_1 = \frac MN$ is the mass of a single helix turn. (A more general
version replaces $N$ by $A/\pi R^2$.) It should be added that $T$ is a function of
length because it depends on the mass.

The criterion \pref{eq:switchi} along its range of validity will be demonstrated
numerically in the following section.

\subsection{\label{sec:numr1}Numerical results for single-projectile dynamics}
After the analytic considerations, we resort to the numerical solution of the
EOM in order to
investigate the situation in which the rotational invariance is broken by
the potential, \epref{eq:v}, which sets three preferred directions.
First, we inspect the applicability of the SA and then we investigate the
hindered helix subject to the current.
\subsubsection{Limits of the sudden approximation}

\begin{figure}
\includegraphics[width=0.95\columnwidth]{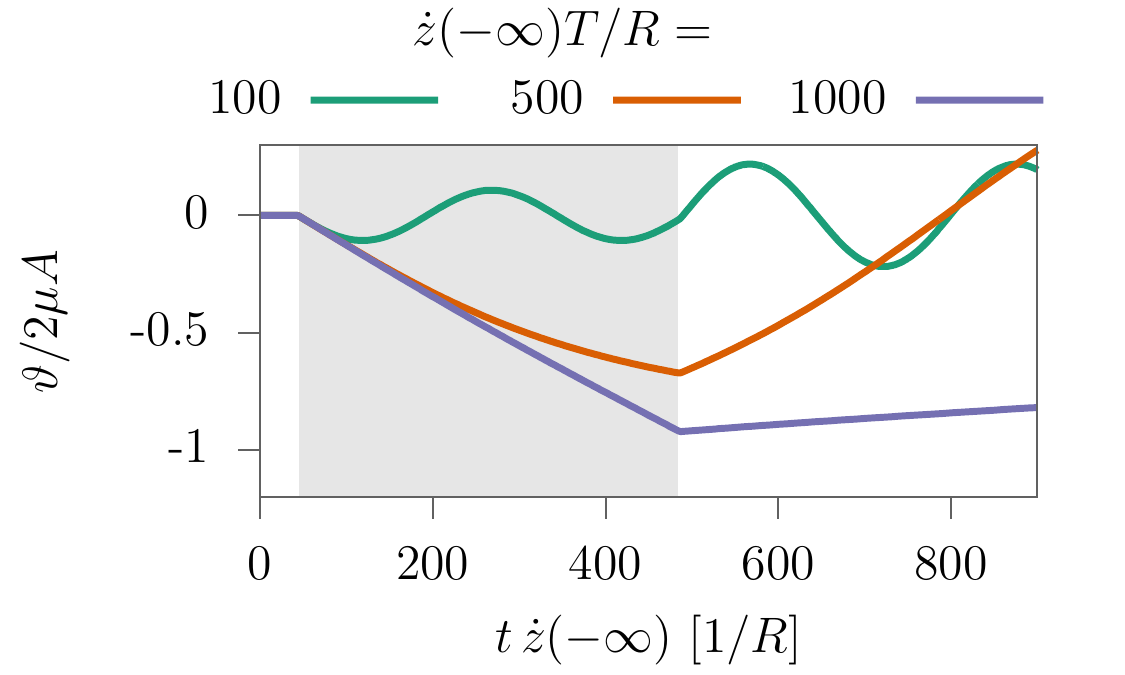}
\caption{\label{fig:1shot}
    Impact of the passage of a single particle
    on the angle of the 50-helix. The shaded region denotes the time interval
    when $\rho$ is nonzero, \ie, the particle is in the helix.
    The traces are parameterized by the impact velocity.
    The time is rescaled by the velocity in order to match the traces together.
    The angle is rescaled by factors from the \epref{eq:turn}, whence -1
    indicates the angle in the sudden limit.
    Other parameters are $\mu = m/M = 0.004$, $\delta N = 1$, and the potential
    of \epref{eq:v}.}
\end{figure}

Figure~\ref{fig:1shot} shows the time evolution of $\vartheta$ during
and after a collision with a particle for three different impact velocities.
The initial condition for the helix was $\vartheta=\dot\vartheta = 0$ and
the particle was put at the entrance of the helix with impact velocity
$\dot z(-\infty)$.

At the beginning of the collision, for very short times,
the three curves lie on top of each other.
This is because at these times the restoring torque $-V'(\vartheta)$ is not
very effective. The condition that the passage time is much smaller than
$T$ is fulfilled at the beginning for all traces. For $\dzitr = 1000$
the helix turns with an almost constant velocity when the particle is in
and the total angle reaches the value from the SA.
For $\dzitr = 500$ the restoring torque markedly bends the curve towards
the potential minimum. The slowest projectile (green curve) gives the helix
only a small initial velocity. When the projectile is in the body of the helix, the helix
oscillates. Here, SA is not applicable, except for the very short times.
The collapse of the SA predicted by the Eqs.~(\ref{eq:condt},\ref{eq:condt2}) is
$\dzitr \approx 314$, consistent with the Figure~\ref{fig:1shot}.

After the collision, the
helix performs either a bound motion around the potential minimum or
its motion is unbound.
In Supplementary Fig.~\ref{fig:n} we plot the 
critical parameters $m/M$, $\dot z(-\infty)$ and $N$.

\subsection{Helix under a current}
The threshold ratio $m/M_1 = 1/6$ is too high to be achieved in molecules under STM,
but we can make it
more favorable if we consider the helix under a particle current $I$, as
the formula \epref{eq:switchi} suggests.

\begin{figure}
\centering
\includegraphics[width=\columnwidth]{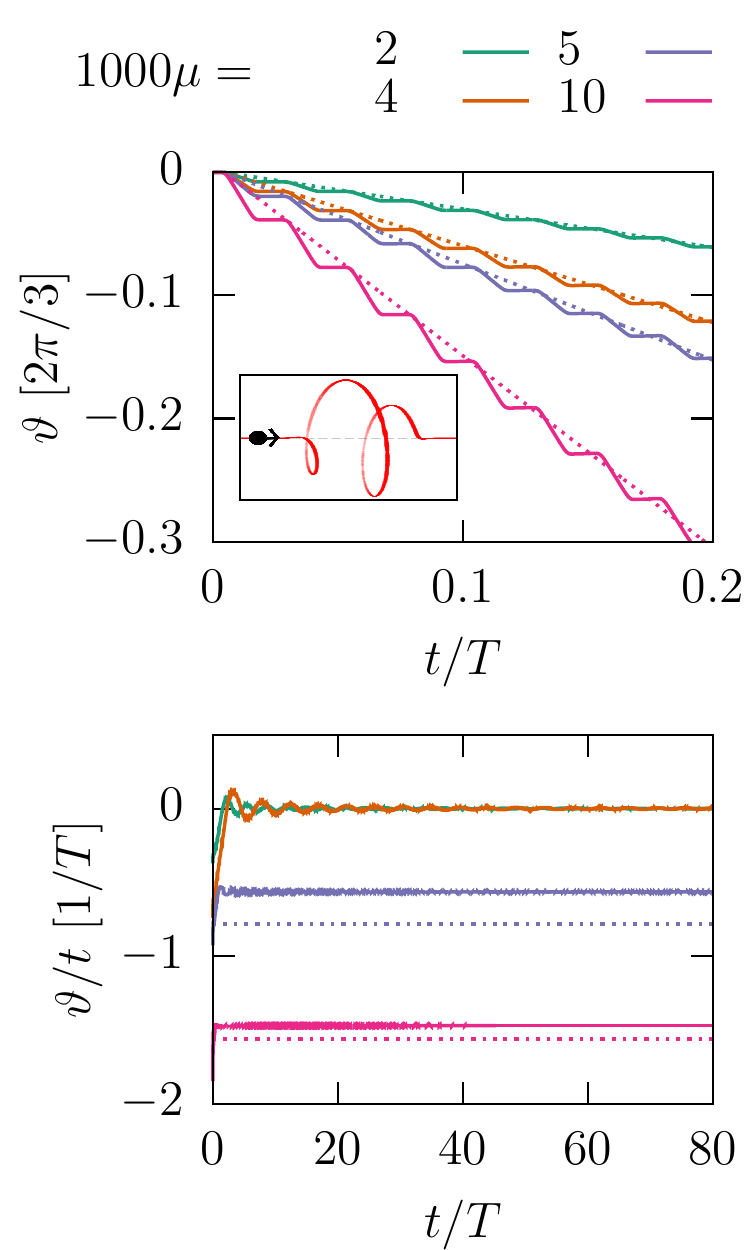}
\caption{\label{fig:traces} 
Time evolution of the angle of the path $\vartheta$ under a current $IT=40$,
for different mass ratios $\mu = m/M$ and $\dzitr = 1000$.
The path is an $N$-helix with $N=1.5, \delta N = 1$ (depicted in the inset).
Numerical results (solid lines) are complemented by
linear evolution from the SA (dotted lines).
The steep parts of the saw-tooth profile, visible in the top panel,
are in the intervals when particle flies through the helix.
The bottom panel presents $\vartheta/t$ for long times, showing directed motion
for large enough $\mu$.
}
\end{figure}

Figure~\ref{fig:traces} shows the evolution of the angle when the 1.5-helix
is under a current. The traces contain tiny sequential steps, which are more
pronounced for large $\mu$. These are the angular boosts produced by the collisions.
The plot also shows a comparison with a straight line obtained from the 
SA:  $\vartheta_\text{SA}(t) = -tI_m 2A/\Theta$ (smoothed over time).
The deviation is caused by the restoring torque $-V'(\vartheta)$, which
counteracts the boosts. This counter-effect can result in a \textit{bound (oscillatory)
motion} or an \textit{unbound directed motion} with a constant
average angular velocity $\langle \dot \vartheta \rangle$ and a small oscillatory component.

\begin{figure}
\includegraphics[width=\columnwidth]{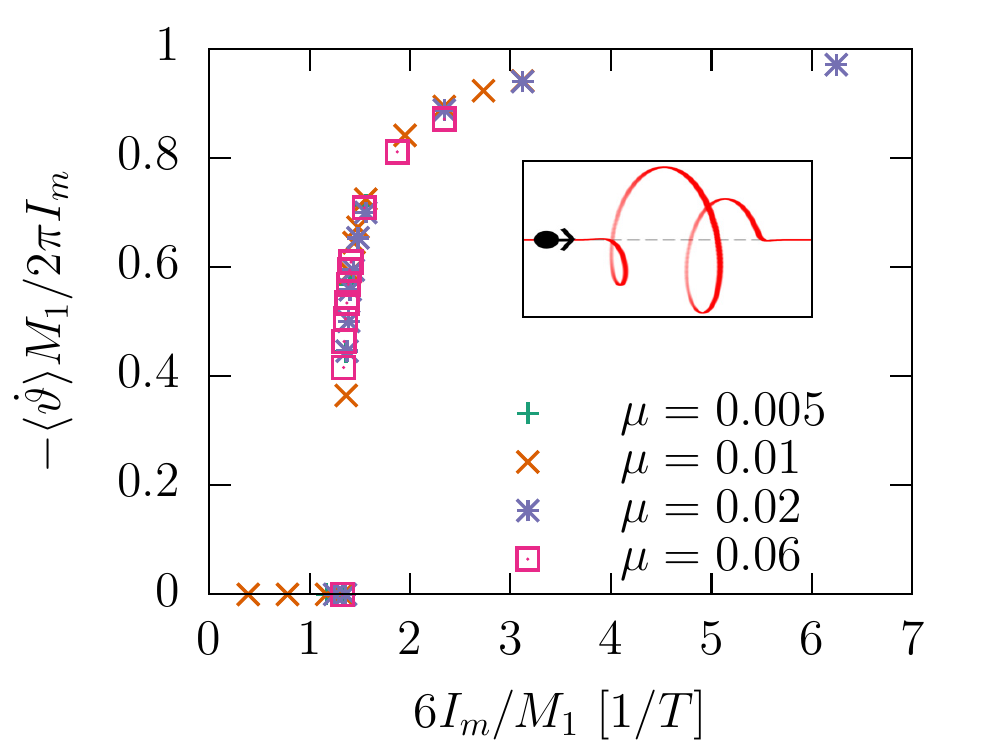}
\caption{\label{fig:dtheta} Dependence of the average angular velocity
of the helix under a mass current $I_m = mI$ for different mass ratios
$\mu = m/M$. $M_1$ is the mass of a single helix turn; the impact
velocity is $\dzitr = 1000$. The inset shows that the particle upon entering
the helix starts revolving counterclockwise when seen from the opposite helix end. Therefore, the helix turns clockwise
to compensate the angular momentum. The current therefore causes a constant
negative $-\langle\dot\vartheta\rangle$.
}
\end{figure}

How does $\langle \dot \vartheta \rangle$ depend on the current?
\epref{eq:switchi} suggest a critical behavior as $I_m$ increases.
For large currents, a linear relation $\langle \dot \vartheta \rangle
\propto - I_m$ is expected, because each particle causes
an angular boost. Fig.~\ref{fig:dtheta}
shows the dependence of the velocity on the mass current for different
mass ratios in the fast impact limit. The data points collapse
on a single universal curve.


This plot fully encapsulates the mass dependence.
It also has the length dependence: as a function of length, only $T$ is expected
to change via the linear increase of $M = M_1N$.
Provided the impact velocity is fast enough, the universal curve has negligible
velocity dependence.
Supplementary Figure~\ref{fig:velocity_dep} shows that for smaller velocities,
the threshold $I_m$ shifts to higher values.
The limit $\Delta V\rightarrow 0^+$ is also of interest: It implies $T\rightarrow\infty$,
and henceforth vanishing threshold $I_m$.

\subsection{\label{sec:ghose}Directed motion of a helix with an open end}
\paragraph{Angular momentum transfer.}
In a scanning tunneling setup, the condition \pref{eq:zerorho} is not always
realized, for example, when the tip of the microscope does not bind to the molecule.
In our theoretical framework, this situation is represented by a path parameterized by 
$s \in (-\infty, s_\mathrm F)$. At the initial point $\rho(s=-\infty) = 0$,
but at the final point $\rho(s_\mathrm F) := s_\mathrm F > 0$.
Thus, as the particle leaves the path at $s_\mathrm F$, it transfers angular momentum
to the path, see \epref{eq:am}. Consequently, the collision causes a
boost both in $\vartheta$ and $\dot\vartheta$.

As long as the restoring torque can be neglected, in the SA we can
obtain the angular momentum boost by combining energy and angular momentum 
conservation laws,
\begin{equation}
\label{eq:ghose}
\Theta \Delta \dot\vartheta = -\rho_\mathrm F m \dot z(-\infty) + \mathcal O(\frac mM)^2.
\end{equation}
The first term assumes that the velocity of the outgoing particle
equals the impact velocity. This velocity must be corrected due to energy
transfer, which yields a term in the second order in $\mu$.

\paragraph{Switching in the SA.}
The condition for switching is that the energy gain of the
path, \epref{eq:energy}, must overcome the potential barrier. In the limit of large
velocities the kinetic term (due to angular momentum boosts) dominates
over the potential gain via angular boosts
and the condition becomes $\Delta V = \half \Theta (\Delta\dot\vartheta)^2$.
For a single particle, the switching condition reads $\sqrt 2 =
\rho_\mathrm F \mu \dot z(-\infty) T/R^2$.

Under a current, the velocity boosts can be added sequentially
and the condition becomes
\begin{equation}
\sqrt 2 =
     \rho_\mathrm F I_m \dot z(-\infty) \frac{T^2}{MR^2}
 = \frac{ \rho_\mathrm F I_p}{\Delta V}
\end{equation}
where we introduced the incident momentum current
$I_p = m \dot z(-\infty) I$. The nominator of the fraction
on the right side can be interpreted approximately as the outgoing
angular momentum current, in view of the expansion in \epref{eq:ghose}.

\begin{figure}
\includegraphics[width=\columnwidth]{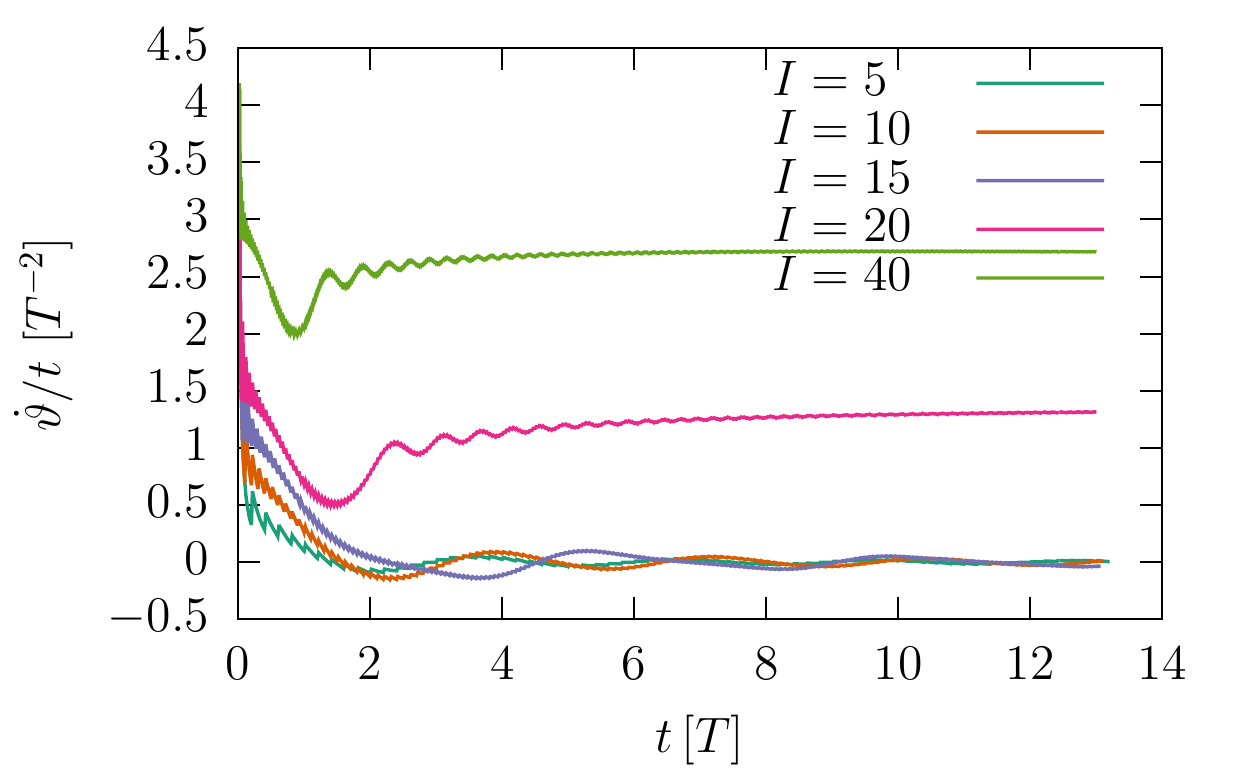}
\caption{\label{fig:ghose_traces} $\rho_\mathrm F = R$. Time evolution of the angular
velocity for a path with $\rho_\mathrm F = R$, where the angular momentum boosts
dominate. As a function of $\mu$, there is a transition from a bound motion
to an unbound motion, with a non-constant velocity.}
\end{figure}

Numerical simulations confirm the threshold behavior, Fig.~\ref{fig:ghose_traces}.
Below the threshold, the path is bound to the potential minimum. Above the
transition the path is accelerated, possibly non-uniformly.

\begin{figure}
\includegraphics[width=\columnwidth]{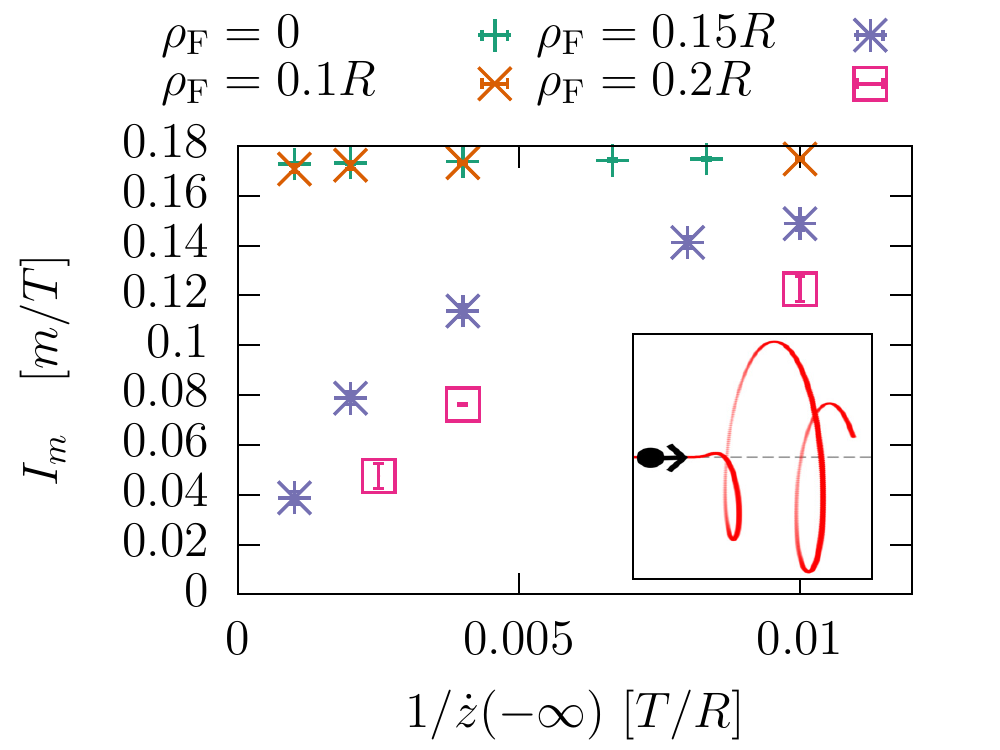}
\caption{\label{fig:threshold_current} Threshold mass current as a function of the
impact velocity for different values of the exit radius $\rho_\mathrm F$. $\mu = 0.05$}
\end{figure}

\begin{figure}[ht]
\includegraphics[width=.95\columnwidth]{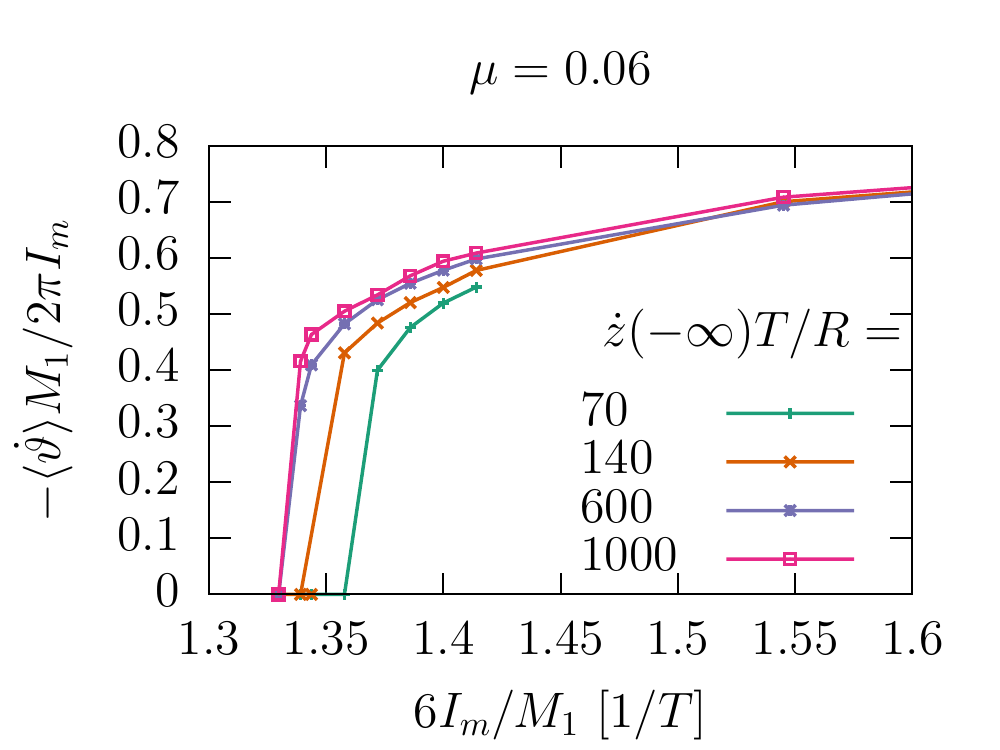}
\caption{\label{fig:velocity_dep} Velocity dependence of average angular velocity of the
path subjected to a mass current $I_m$. For lower velocities the threshold current
increases, marking a departure from the SA.}
\end{figure}

In the next step we investigate the dependence of the threshold current. For a fixed
mass ratio, Fig.~\ref{fig:threshold_current} shows the threshold mass current.
There are two regimes covered in that plot:
\begin{itemize}
\item $\rho_\mathrm F = 0$: the helix is not accelerated. The threshold $I_m$ 
      depends on the impact velocity very weakly in the given range.
      Actually, it increases with decreasing impact velocity (see Fig.~\ref{fig:velocity_dep}. This is the
      mechanism of \textit{angular boosts} studied in the previous section.
\item For nonzero $\rho_\mathrm F$, the collision causes a net torque,
      the helix always accelerates.
      In the limit of large impact velocities the switching due to
      \textit{angular momentum boosts} overtakes and the threshold $I_m$ drops
      inversely proportional to $\dot z(-\infty)$. This regime is the
      familiar \textit{garden hose} effect.
\end{itemize}

\begin{figure}
\includegraphics[width=\columnwidth]{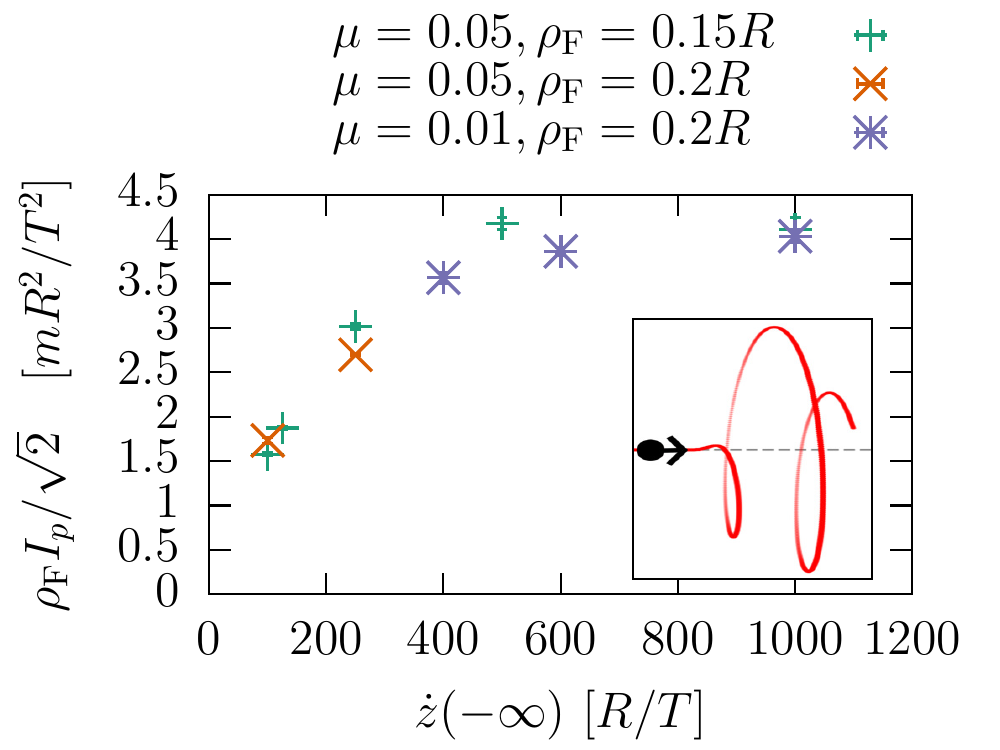}
\caption{\label{fig:momentum_current} Velocity dependence of the threshold
momentum current $I_p = m \dot z(-\infty) I$. The large - velocity limit is
dominated by the garden hose effect -- \textit{angular momentum boosts}.}
\end{figure}

To take a closer look at the mechanism of \textit{angular momentum boosts}, we
plot the threshold momentum current $I_p$ for different values of
$\rho_\mathrm F$ and $\mu$ in Fig.~\ref{fig:momentum_current}.
The data collapses on a single curve, which saturates in the large impact velocity
limit.

\section{Discussion and Outlook}
Our results are valid when thermal fluctuations are small, \ie
$k_\mathrm BT \ll \Delta V$. To account for fluctuations,
it is customary to apply the Langevin equation for $\vartheta$
equipped with stochastic torques and a deterministic torque, the latter
driving the directed rotation \cite{Hangi2009}. Our approach predicts the 
detailed form of the
deterministic torque as it follows from the passage of
the particle through a chiral path with $\rho_\mathrm F \sim R$.
When $\rho_\mathrm F = 0$, the effect of the passage (collision) is
to boost the angle. In the stochastic equation, such a single-particle process
can be accounted for by the torque of the form
\[
F(\vartheta, \dot\vartheta)\frac{\mathrm d}{\mathrm dt}\delta(t)
\]
in the limit of short collision times. The derivative of a delta function expresses a torque pulse
that is immediately cancelled by a pulse of an opposite sign, thus
generating no net $\dot\vartheta$ but a boost in $\vartheta$.
The function $F(\vartheta, \dot\vartheta)$ follows from our methodology straightforwardly.

We have focused largely on the conditions of a directed
rotation. To implement an efficient switch, more conditions
need to be fulfilled. First, the rotor's velocity must be
attenuated in order for the rotor to settle in the nearest
potential minimum. Second, the current must flow in controlled short pulse. The optimal parameter regime can be sought using
the EOM \eprefs{eq:eom1}{eq:eom2} and it
is beyond the scope of the present work.  We expect that the threshold current will vanish in a strongly overdamped limit and a linear response of $\dot \vartheta$ is expected.

Quantum effects are responsible for rich transport phenomenology of
molecular junctions \cite{Evers2020}. Here we pause to discuss
quantum effects related to the electronic degrees of freedom, assuming
that the quantization levels of the rotational motion fall below the
working temperature. Rotation only happens via inelastic electron
 tunneling. Importantly, each single
electron scattering event must obey fundamental conservation laws;
therefore, the principles outlined in this manuscript will survive
in the quantum limit. Two quantum aspects are significant in this context:
(1) the electron transport process is stochastic, allowing
for transmission and reflection at the same time. Particle reflection off
the helix can not induce any rotation, unless the following
effect is considered.
(2) electrons carry spin angular momentum, which couples with
the orbital momentum by spin-orbit interaction (SOI). It was
proposed that SOI is responsible for spin selectivity in helical
wires \cite{Evers2022}. In relation to the topic of molecular rotors,
the question of how could the spin degree of
freedom induce mechanical rotation is an open one.

The results presented here demonstrate a directed rotation without
any momentum transfer in a molecular \emph{rotor}. Such devices
can rotate when under particle current, but they can not do work, because
each electron boosts the angle but not the frequency. Although
they can not operate as \emph{motors}, these rotors can serve in
nanoscale information storage and processing. The information readout
can be performed in linear response (under the threshold current).
A small symmetry breaking is needed in order to discriminate between the three states.

\section{Conclusions}
Summarizing, we have investigated the classical dynamics of a molecular rotor under
a particle current. The molecule was modelled by a massive path that has a rotational
degree of freedom. Our approach expresses the impact of
a single collision on the rotor in a way that stems explicitly from the
(chiral) geometry of the rotor.

When the particles do not carry (or take) any angular momentum, rotation is
possible via \textit{angular boosts}. If the rotation is hindered by a potential barrier
$\Delta V$, the requirement that
the incident particles carry enough energy is not sufficient for switching.
Instead, a stricter requirement
that the boosts be sufficiently fast and dense in time applies, \epref{eq:switchi}.

When the particles are allowed to transfer angular momentum, we predict
a crossover from the regime of \textit{angular boosts} to the regime
of \textit{angular momentum boosts}.

\begin{acknowledgments}
F.\ Evers acknowledges funding from the Deutsche Forschungsgemeinschaft
(DFG, German Research Foundation) through CRC 1277 subprojects A03 and B01. 
R.\ Korytár acknowledges the Czech Science 
foundation (project no. 22-22419S) and support of the Ministry of Education, Youth and
Sports of the Czech Republic through the e-INFRA CZ (ID:90140).
\end{acknowledgments}

\section*{Appendix: Coordinates of an $N$-helix}
We introduce a path definition
\begin{align}
\phi(s) &= s\\
\rho(s) &= \bigl\{\erfc{ 2(s/\pi -N)/\delta N}  -\\
        &{\phantom{= \bigl\{}}          \erfc{ 2(s/\pi +N)/\delta N} \bigr\}/2\\
z(s)    &=  s
\end{align}
describing a helix with $N$ turns, whose radius goes to zero smoothly at its
both ends within a distance proportional to $\delta N$, see Fig.~\ref{fig:scheme}(b).
The smooth onset is achieved by employing the complementary error functions, erfc.
In the above definition, we adopt as a unit of length the maximum radius $R$.

This path will be employed for $s\in (-\infty,s_\mathrm F)$. When $s_\mathrm F=\infty$,
condition \pref{eq:zerorho} is fulfilled.
Setting as finite $s_\mathrm F$ provides a path with an open end, 
when the particle exits the path at a finite $\rho$.

\section*{Appendix: Equations of motion}
\subsection{EOM from the Lagrangian}
The equations of motion (EOM) derived from a Lagrangian $\mcL(\vartheta, \dot\vartheta, s,\dot s)$
by the principle of least action \cite{Landau} read
\begin{align}
\frac {\mathrm d}{\mathrm dt}\left(\frac{\partial \mcL}{\partial \dot\vartheta}\right)
&= \frac{\partial \mcL}{\partial \vartheta}\\
\frac {\mathrm d}{\mathrm dt}\left(\frac{\partial \mcL}{\partial \dot s}\right)
&= \frac{\partial \mcL}{\partial s}.
\end{align}

Inserting \epref{eq:lagrangian}, the EOM take the form
\begin{multline}
\label{eq:eom1}
m\varrho^2 \phio'\,\ddot s + m(\rho^2 \phio')'\,\dot s^2 + (2m\rho\rho')\,\dot s \dot\vartheta\, +\Thetarho\, \ddot\vartheta =\\
= - V'(\vartheta)
- \gamma\Theta\,\dot\vartheta
\end{multline}

\begin{multline}
\label{eq:eom2}
\left(\rho'^2 +\rho^2\phio'^2 + z'^2\right)\, \ddot s\, +\frac 12 \left(\rho'^2 +\rho^2\phio'^2 + z'^2\right)'\,\dot s^2
- \\ 
 - \rho\rho'\, \dot\vartheta ^2 +\,\rho^2 \phio'\,\ddot\vartheta = 0
\end{multline}
The first EOM delivers the equation for a rotor in the limit $m=0$. We have added a phenomenological damping term $-\gamma\Theta\,\dot\vartheta$ to the EOM (which does not follow from the conservative Lagrangian formalism). The damping
term is zero in all numerical results of this work unless stated explicitly.
When $V=\gamma =0$,
the equation expresses angular momentum conservation. The second
equation describes the constrained particle dynamics if $\vartheta = \text{const.}$
Notice that the mass $m$ drops out, because the particle experiences
inertial forces only.


\subsection{Transformation to dimension-less variables}
\subsubsection{EOM of the rotor}
Substituting $\tilde t = t / T$ in Eq.~(\ref{eq:eom1})
renders the first EOM dimension-less,
\begin{multline}
    (1 + \frac mM \tilde{\rho}^2)\, \ddot\vartheta\, +
    \frac mM\tilde{\varrho}^2 \phio'\,\ddot s + \frac mM (\tilde{\rho}^2 \phio')'\,\dot s^2 + 2\frac mM \tilde{\rho}\tilde{\rho}'\,\dot s \dot\vartheta =\\
= - V'(\vartheta)/\Delta V - \tilde{\gamma}\,\dot\vartheta
\end{multline}
where now the dots indicate differentiation with respect to $\tilde t$ and
we defined $\tilde{\rho} := \rho / R$, which is a quantity of the order
of unity, and $\tilde{\gamma} := \gamma T$ is a dimensionless damping
rate.
Notice the appearance of the small parameter $\frac mM $,
which, however, is here often multiplied by the large velocity $\dot s$.

\subsubsection{EOM for the particle}
The substitution of $\tilde{t}$ leaves the second EOM in the form
\begin{multline}
\label{eom2d}
\left(\tilde{\rho}'^2 +\tilde{\rho}^2\phio'^2 + \tilde{z}'^2\right)\, \ddot s\, +\frac 12 \left(\tilde{\rho}'^2 +\tilde{\rho}^2\phio'^2 + \tilde{z}'^2\right)'\,\dot s^2
- \\ 
 - \tilde{\rho}\tilde{\rho}'\, \dot\vartheta ^2 +\,\tilde{\rho}^2 \phio'\,\ddot\vartheta = 0,
\end{multline}
where we introduced $\tilde z = z/R$ and the dot indicates differentiation w.r. to $\tilde t$.
At the entry point, $s=0$, the velocity
of the particle equals $\mathrm d z(0)/\mathrm dt$. 
The $\mathrm d{\tilde{z}}(0)/\mathrm d t$ means the inverse
dwell time $\delta\tau$. We shall assume that $\delta\tau = 10^{-2}T$ and henceforth $\mathrm d \tilde{z}(0)/\mathrm d\tilde{t}\approx 10^{2}$.


\section*{Appendix: Peripheral numerical results}

\begin{figure}[ht]
\includegraphics[width=0.95\columnwidth]{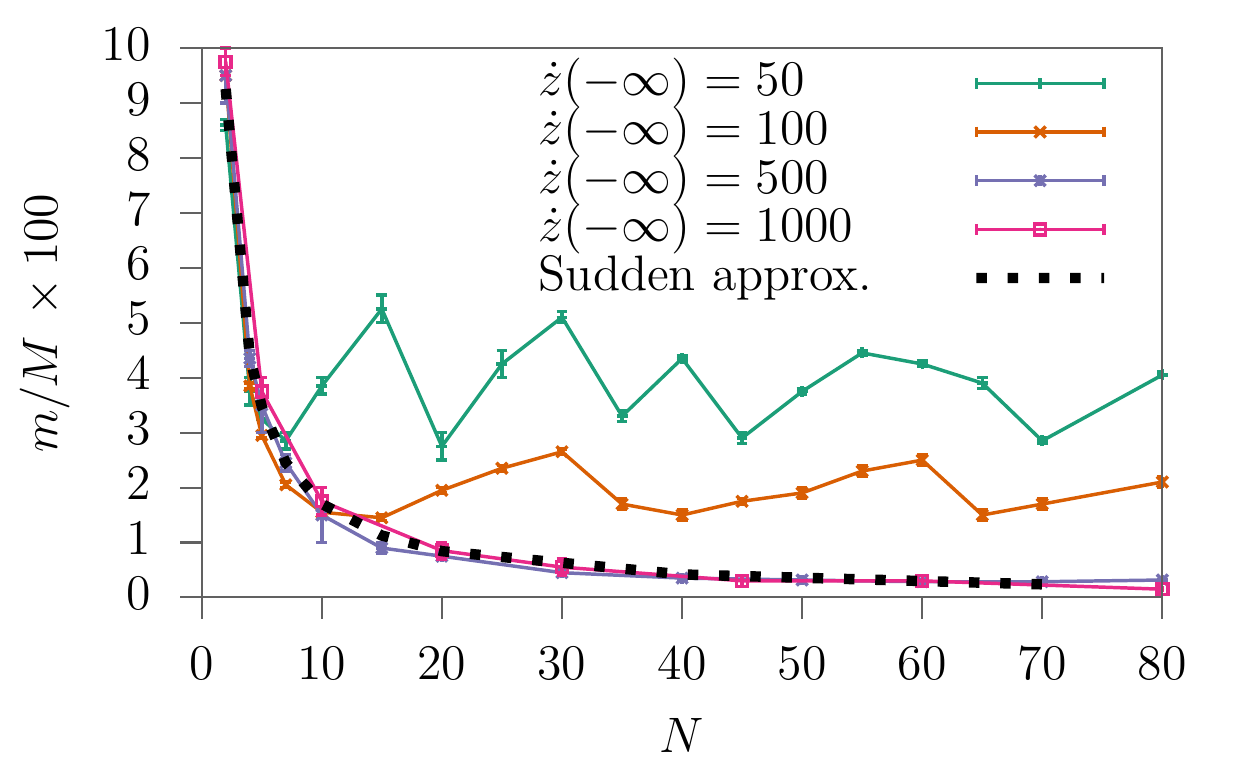}
\caption{\label{fig:n} Threshold mass ratio $m/M$ for an unbound directed motion
as a function of helix length $N$ ($\delta N=1$) and the impact velocity. The dotted line
is the threshold according to the SA, \epref{eq:sudden1}, which coincided with the numerics
if the time particle spends in the helix is short.
}
\end{figure}

\clearpage

\bibliography{references}

\begin{thebibliography}{19}%
\makeatletter
\providecommand \@ifxundefined [1]{%
 \@ifx{#1\undefined}
}%
\providecommand \@ifnum [1]{%
 \ifnum #1\expandafter \@firstoftwo
 \else \expandafter \@secondoftwo
 \fi
}%
\providecommand \@ifx [1]{%
 \ifx #1\expandafter \@firstoftwo
 \else \expandafter \@secondoftwo
 \fi
}%
\providecommand \natexlab [1]{#1}%
\providecommand \enquote  [1]{``#1''}%
\providecommand \bibnamefont  [1]{#1}%
\providecommand \bibfnamefont [1]{#1}%
\providecommand \citenamefont [1]{#1}%
\providecommand \href@noop [0]{\@secondoftwo}%
\providecommand \href [0]{\begingroup \@sanitize@url \@href}%
\providecommand \@href[1]{\@@startlink{#1}\@@href}%
\providecommand \@@href[1]{\endgroup#1\@@endlink}%
\providecommand \@sanitize@url [0]{\catcode `\\12\catcode `\$12\catcode
  `\&12\catcode `\#12\catcode `\^12\catcode `\_12\catcode `\%12\relax}%
\providecommand \@@startlink[1]{}%
\providecommand \@@endlink[0]{}%
\providecommand \url  [0]{\begingroup\@sanitize@url \@url }%
\providecommand \@url [1]{\endgroup\@href {#1}{\urlprefix }}%
\providecommand \urlprefix  [0]{URL }%
\providecommand \Eprint [0]{\href }%
\providecommand \doibase [0]{https://doi.org/}%
\providecommand \selectlanguage [0]{\@gobble}%
\providecommand \bibinfo  [0]{\@secondoftwo}%
\providecommand \bibfield  [0]{\@secondoftwo}%
\providecommand \translation [1]{[#1]}%
\providecommand \BibitemOpen [0]{}%
\providecommand \bibitemStop [0]{}%
\providecommand \bibitemNoStop [0]{.\EOS\space}%
\providecommand \EOS [0]{\spacefactor3000\relax}%
\providecommand \BibitemShut  [1]{\csname bibitem#1\endcsname}%
\let\auto@bib@innerbib\@empty
\bibitem [{\citenamefont {Lensen}\ and\ \citenamefont
  {Elemans}(2012)}]{Lensen2012}%
  \BibitemOpen
  \bibfield  {author} {\bibinfo {author} {\bibfnamefont {D.}~\bibnamefont
  {Lensen}}\ and\ \bibinfo {author} {\bibfnamefont {J.~A.}\ \bibnamefont
  {Elemans}},\ }\href@noop {} {\bibfield  {journal} {\bibinfo  {journal} {Soft
  Matter}\ }\textbf {\bibinfo {volume} {8}},\ \bibinfo {pages} {9053} (\bibinfo
  {year} {2012})}\BibitemShut {NoStop}%
\bibitem [{\citenamefont {Perera}\ \emph {et~al.}(2013)\citenamefont {Perera},
  \citenamefont {Ample}, \citenamefont {Kersell}, \citenamefont {Zhang},
  \citenamefont {Vives}, \citenamefont {Echeverria}, \citenamefont {Grisolia},
  \citenamefont {Rapenne}, \citenamefont {Joachim},\ and\ \citenamefont
  {Hla}}]{Perera2013}%
  \BibitemOpen
  \bibfield  {author} {\bibinfo {author} {\bibfnamefont {U.~G.~E.}\
  \bibnamefont {Perera}}, \bibinfo {author} {\bibfnamefont {F.}~\bibnamefont
  {Ample}}, \bibinfo {author} {\bibfnamefont {H.}~\bibnamefont {Kersell}},
  \bibinfo {author} {\bibfnamefont {Y.}~\bibnamefont {Zhang}}, \bibinfo
  {author} {\bibfnamefont {G.}~\bibnamefont {Vives}}, \bibinfo {author}
  {\bibfnamefont {J.}~\bibnamefont {Echeverria}}, \bibinfo {author}
  {\bibfnamefont {M.}~\bibnamefont {Grisolia}}, \bibinfo {author}
  {\bibfnamefont {G.}~\bibnamefont {Rapenne}}, \bibinfo {author} {\bibfnamefont
  {C.}~\bibnamefont {Joachim}},\ and\ \bibinfo {author} {\bibfnamefont
  {S.}~\bibnamefont {Hla}},\ }\href@noop {} {\bibfield  {journal} {\bibinfo
  {journal} {Nature Nanotechnology}\ }\textbf {\bibinfo {volume} {8}},\
  \bibinfo {pages} {46} (\bibinfo {year} {2013})}\BibitemShut {NoStop}%
\bibitem [{\citenamefont {Schaffert}\ \emph {et~al.}(2013)\citenamefont
  {Schaffert}, \citenamefont {Cottin}, \citenamefont {Sonntag}, \citenamefont
  {Karacuban}, \citenamefont {Bobisch}, \citenamefont {Lorente}, \citenamefont
  {Gauyacq},\ and\ \citenamefont {M{\"o}ller}}]{Schaffert2013}%
  \BibitemOpen
  \bibfield  {author} {\bibinfo {author} {\bibfnamefont {J.}~\bibnamefont
  {Schaffert}}, \bibinfo {author} {\bibfnamefont {M.~C.}\ \bibnamefont
  {Cottin}}, \bibinfo {author} {\bibfnamefont {A.}~\bibnamefont {Sonntag}},
  \bibinfo {author} {\bibfnamefont {H.}~\bibnamefont {Karacuban}}, \bibinfo
  {author} {\bibfnamefont {C.~A.}\ \bibnamefont {Bobisch}}, \bibinfo {author}
  {\bibfnamefont {N.}~\bibnamefont {Lorente}}, \bibinfo {author} {\bibfnamefont
  {J.-P.}\ \bibnamefont {Gauyacq}},\ and\ \bibinfo {author} {\bibfnamefont
  {R.}~\bibnamefont {M{\"o}ller}},\ }\href@noop {} {\bibfield  {journal}
  {\bibinfo  {journal} {Nature Mater.}\ }\textbf {\bibinfo {volume} {12}},\
  \bibinfo {pages} {223} (\bibinfo {year} {2013})}\BibitemShut {NoStop}%
\bibitem [{\citenamefont {Ohmann}\ \emph {et~al.}(2015)\citenamefont {Ohmann},
  \citenamefont {Meyer}, \citenamefont {Nickel}, \citenamefont {Echeverria},
  \citenamefont {Grisolia}, \citenamefont {Joachim}, \citenamefont {Moresco},\
  and\ \citenamefont {Cuniberti}}]{Ohmann2015}%
  \BibitemOpen
  \bibfield  {author} {\bibinfo {author} {\bibfnamefont {R.}~\bibnamefont
  {Ohmann}}, \bibinfo {author} {\bibfnamefont {J.}~\bibnamefont {Meyer}},
  \bibinfo {author} {\bibfnamefont {A.}~\bibnamefont {Nickel}}, \bibinfo
  {author} {\bibfnamefont {J.}~\bibnamefont {Echeverria}}, \bibinfo {author}
  {\bibfnamefont {M.}~\bibnamefont {Grisolia}}, \bibinfo {author}
  {\bibfnamefont {C.}~\bibnamefont {Joachim}}, \bibinfo {author} {\bibfnamefont
  {F.}~\bibnamefont {Moresco}},\ and\ \bibinfo {author} {\bibfnamefont
  {G.}~\bibnamefont {Cuniberti}},\ }\href@noop {} {\bibfield  {journal}
  {\bibinfo  {journal} {ACS Nano}\ }\textbf {\bibinfo {volume} {9}},\ \bibinfo
  {pages} {8394} (\bibinfo {year} {2015})}\BibitemShut {NoStop}%
\bibitem [{\citenamefont {Simpson}\ \emph {et~al.}(2019)\citenamefont
  {Simpson}, \citenamefont {Garc{\'\i}a-L{\'o}pez}, \citenamefont
  {Daniel~Boese}, \citenamefont {Tour},\ and\ \citenamefont
  {Grill}}]{Simpson2019}%
  \BibitemOpen
  \bibfield  {author} {\bibinfo {author} {\bibfnamefont {G.~J.}\ \bibnamefont
  {Simpson}}, \bibinfo {author} {\bibfnamefont {V.}~\bibnamefont
  {Garc{\'\i}a-L{\'o}pez}}, \bibinfo {author} {\bibfnamefont {A.}~\bibnamefont
  {Daniel~Boese}}, \bibinfo {author} {\bibfnamefont {J.~M.}\ \bibnamefont
  {Tour}},\ and\ \bibinfo {author} {\bibfnamefont {L.}~\bibnamefont {Grill}},\
  }\href@noop {} {\bibfield  {journal} {\bibinfo  {journal} {Nature Commun.}\
  }\textbf {\bibinfo {volume} {10}},\ \bibinfo {pages} {1} (\bibinfo {year}
  {2019})}\BibitemShut {NoStop}%
\bibitem [{\citenamefont {Jasper-Toennies}\ \emph {et~al.}(2020)\citenamefont
  {Jasper-Toennies}, \citenamefont {Gruber}, \citenamefont {Johannsen},
  \citenamefont {Frederiksen}, \citenamefont {Garcia-Lekue}, \citenamefont
  {Jäkel}, \citenamefont {Roehricht}, \citenamefont {Herges},\ and\
  \citenamefont {Berndt}}]{JasperTonnies2020}%
  \BibitemOpen
  \bibfield  {author} {\bibinfo {author} {\bibfnamefont {T.}~\bibnamefont
  {Jasper-Toennies}}, \bibinfo {author} {\bibfnamefont {M.}~\bibnamefont
  {Gruber}}, \bibinfo {author} {\bibfnamefont {S.}~\bibnamefont {Johannsen}},
  \bibinfo {author} {\bibfnamefont {T.}~\bibnamefont {Frederiksen}}, \bibinfo
  {author} {\bibfnamefont {A.}~\bibnamefont {Garcia-Lekue}}, \bibinfo {author}
  {\bibfnamefont {T.}~\bibnamefont {Jäkel}}, \bibinfo {author} {\bibfnamefont
  {F.}~\bibnamefont {Roehricht}}, \bibinfo {author} {\bibfnamefont
  {R.}~\bibnamefont {Herges}},\ and\ \bibinfo {author} {\bibfnamefont
  {R.}~\bibnamefont {Berndt}},\ }\href@noop {} {\bibfield  {journal} {\bibinfo
  {journal} {ACS Nano}\ }\textbf {\bibinfo {volume} {14}},\ \bibinfo {pages}
  {3907} (\bibinfo {year} {2020})}\BibitemShut {NoStop}%
\bibitem [{\citenamefont {Wu}\ \emph {et~al.}(2020)\citenamefont {Wu},
  \citenamefont {Liu}, \citenamefont {Zhang}, \citenamefont {Wang},
  \citenamefont {Shen}, \citenamefont {Li}, \citenamefont {Berndt},
  \citenamefont {Hou},\ and\ \citenamefont {Wang}}]{Wu2020}%
  \BibitemOpen
  \bibfield  {author} {\bibinfo {author} {\bibfnamefont {T.}~\bibnamefont
  {Wu}}, \bibinfo {author} {\bibfnamefont {L.}~\bibnamefont {Liu}}, \bibinfo
  {author} {\bibfnamefont {Y.}~\bibnamefont {Zhang}}, \bibinfo {author}
  {\bibfnamefont {Y.}~\bibnamefont {Wang}}, \bibinfo {author} {\bibfnamefont
  {Z.}~\bibnamefont {Shen}}, \bibinfo {author} {\bibfnamefont {N.}~\bibnamefont
  {Li}}, \bibinfo {author} {\bibfnamefont {R.}~\bibnamefont {Berndt}}, \bibinfo
  {author} {\bibfnamefont {S.}~\bibnamefont {Hou}},\ and\ \bibinfo {author}
  {\bibfnamefont {Y.}~\bibnamefont {Wang}},\ }\href@noop {} {\bibfield
  {journal} {\bibinfo  {journal} {Chemical Communications}\ }\textbf {\bibinfo
  {volume} {56}},\ \bibinfo {pages} {968} (\bibinfo {year} {2020})}\BibitemShut
  {NoStop}%
\bibitem [{\citenamefont {Eisenhut}\ \emph {et~al.}(2021)\citenamefont
  {Eisenhut}, \citenamefont {K{\"u}hne}, \citenamefont {Monsalve},
  \citenamefont {Srivastava}, \citenamefont {Ryndyk}, \citenamefont
  {Cuniberti}, \citenamefont {Aiboudi}, \citenamefont {Lissel}, \citenamefont
  {Zoba{\v{c}}}, \citenamefont {Robles} \emph {et~al.}}]{Eisenhut2021}%
  \BibitemOpen
  \bibfield  {author} {\bibinfo {author} {\bibfnamefont {F.}~\bibnamefont
  {Eisenhut}}, \bibinfo {author} {\bibfnamefont {T.}~\bibnamefont {K{\"u}hne}},
  \bibinfo {author} {\bibfnamefont {J.}~\bibnamefont {Monsalve}}, \bibinfo
  {author} {\bibfnamefont {S.}~\bibnamefont {Srivastava}}, \bibinfo {author}
  {\bibfnamefont {D.~A.}\ \bibnamefont {Ryndyk}}, \bibinfo {author}
  {\bibfnamefont {G.}~\bibnamefont {Cuniberti}}, \bibinfo {author}
  {\bibfnamefont {O.}~\bibnamefont {Aiboudi}}, \bibinfo {author} {\bibfnamefont
  {F.}~\bibnamefont {Lissel}}, \bibinfo {author} {\bibfnamefont
  {V.}~\bibnamefont {Zoba{\v{c}}}}, \bibinfo {author} {\bibfnamefont
  {R.}~\bibnamefont {Robles}}, \emph {et~al.},\ }\href@noop {} {\bibfield
  {journal} {\bibinfo  {journal} {Nanoscale}\ }\textbf {\bibinfo {volume}
  {13}},\ \bibinfo {pages} {16077} (\bibinfo {year} {2021})}\BibitemShut
  {NoStop}%
\bibitem [{\citenamefont {Ren}\ \emph {et~al.}(2020)\citenamefont {Ren},
  \citenamefont {Freitag}, \citenamefont {Schwermann}, \citenamefont {Bakker},
  \citenamefont {Amirjalayer}, \citenamefont {Rühling}, \citenamefont {Gao},
  \citenamefont {Doltsinis}, \citenamefont {Glorius},\ and\ \citenamefont
  {Fuchs}}]{Ren2020}%
  \BibitemOpen
  \bibfield  {author} {\bibinfo {author} {\bibfnamefont {J.}~\bibnamefont
  {Ren}}, \bibinfo {author} {\bibfnamefont {M.}~\bibnamefont {Freitag}},
  \bibinfo {author} {\bibfnamefont {C.}~\bibnamefont {Schwermann}}, \bibinfo
  {author} {\bibfnamefont {A.}~\bibnamefont {Bakker}}, \bibinfo {author}
  {\bibfnamefont {S.}~\bibnamefont {Amirjalayer}}, \bibinfo {author}
  {\bibfnamefont {A.}~\bibnamefont {Rühling}}, \bibinfo {author}
  {\bibfnamefont {H.-Y.}\ \bibnamefont {Gao}}, \bibinfo {author} {\bibfnamefont
  {N.~L.}\ \bibnamefont {Doltsinis}}, \bibinfo {author} {\bibfnamefont
  {F.}~\bibnamefont {Glorius}},\ and\ \bibinfo {author} {\bibfnamefont
  {H.}~\bibnamefont {Fuchs}},\ }\href@noop {} {\bibfield  {journal} {\bibinfo
  {journal} {Nano Lett.}\ }\textbf {\bibinfo {volume} {20}},\ \bibinfo {pages}
  {5922} (\bibinfo {year} {2020})}\BibitemShut {NoStop}%
\bibitem [{\citenamefont {Stolz}\ \emph {et~al.}(2020)\citenamefont {Stolz},
  \citenamefont {Gr{\"o}ning}, \citenamefont {Prinz}, \citenamefont {Brune},\
  and\ \citenamefont {Widmer}}]{Stolz2020}%
  \BibitemOpen
  \bibfield  {author} {\bibinfo {author} {\bibfnamefont {S.}~\bibnamefont
  {Stolz}}, \bibinfo {author} {\bibfnamefont {O.}~\bibnamefont {Gr{\"o}ning}},
  \bibinfo {author} {\bibfnamefont {J.}~\bibnamefont {Prinz}}, \bibinfo
  {author} {\bibfnamefont {H.}~\bibnamefont {Brune}},\ and\ \bibinfo {author}
  {\bibfnamefont {R.}~\bibnamefont {Widmer}},\ }\href@noop {} {\bibfield
  {journal} {\bibinfo  {journal} {Proc. Natl. Acad. Sci. U.S.A.}\ }\textbf
  {\bibinfo {volume} {117}},\ \bibinfo {pages} {14838} (\bibinfo {year}
  {2020})}\BibitemShut {NoStop}%
\bibitem [{\citenamefont {H\"anggi}\ and\ \citenamefont
  {Marchesoni}(2009)}]{Hangi2009}%
  \BibitemOpen
  \bibfield  {author} {\bibinfo {author} {\bibfnamefont {P.}~\bibnamefont
  {H\"anggi}}\ and\ \bibinfo {author} {\bibfnamefont {F.}~\bibnamefont
  {Marchesoni}},\ }\href {https://doi.org/10.1103/RevModPhys.81.387} {\bibfield
   {journal} {\bibinfo  {journal} {Rev. Mod. Phys.}\ }\textbf {\bibinfo
  {volume} {81}},\ \bibinfo {pages} {387} (\bibinfo {year} {2009})}\BibitemShut
  {NoStop}%
\bibitem [{\citenamefont {Astumian}\ \emph {et~al.}(2016)\citenamefont
  {Astumian}, \citenamefont {Mukherjee},\ and\ \citenamefont
  {Warshel}}]{Astumian2016}%
  \BibitemOpen
  \bibfield  {author} {\bibinfo {author} {\bibfnamefont {R.~D.}\ \bibnamefont
  {Astumian}}, \bibinfo {author} {\bibfnamefont {S.}~\bibnamefont
  {Mukherjee}},\ and\ \bibinfo {author} {\bibfnamefont {A.}~\bibnamefont
  {Warshel}},\ }\href@noop {} {\bibfield  {journal} {\bibinfo  {journal} {Chem.
  Phys. Chem.}\ }\textbf {\bibinfo {volume} {17}},\ \bibinfo {pages} {1719}
  (\bibinfo {year} {2016})}\BibitemShut {NoStop}%
\bibitem [{\citenamefont {Bode}\ \emph {et~al.}(2011)\citenamefont {Bode},
  \citenamefont {Kusminskiy}, \citenamefont {Egger},\ and\ \citenamefont {von
  Oppen}}]{Bode2011}%
  \BibitemOpen
  \bibfield  {author} {\bibinfo {author} {\bibfnamefont {N.}~\bibnamefont
  {Bode}}, \bibinfo {author} {\bibfnamefont {S.~V.}\ \bibnamefont
  {Kusminskiy}}, \bibinfo {author} {\bibfnamefont {R.}~\bibnamefont {Egger}},\
  and\ \bibinfo {author} {\bibfnamefont {F.}~\bibnamefont {von Oppen}},\ }\href
  {https://doi.org/10.1103/PhysRevLett.107.036804} {\bibfield  {journal}
  {\bibinfo  {journal} {Phys. Rev. Lett.}\ }\textbf {\bibinfo {volume} {107}},\
  \bibinfo {pages} {036804} (\bibinfo {year} {2011})}\BibitemShut {NoStop}%
\bibitem [{\citenamefont {Todorov}\ \emph {et~al.}(2014)\citenamefont
  {Todorov}, \citenamefont {Dundas}, \citenamefont {Lü}, \citenamefont
  {Brandbyge},\ and\ \citenamefont {Hedeg{\aa}rd}}]{Todorov2014}%
  \BibitemOpen
  \bibfield  {author} {\bibinfo {author} {\bibfnamefont {T.~N.}\ \bibnamefont
  {Todorov}}, \bibinfo {author} {\bibfnamefont {D.}~\bibnamefont {Dundas}},
  \bibinfo {author} {\bibfnamefont {J.-T.}\ \bibnamefont {Lü}}, \bibinfo
  {author} {\bibfnamefont {M.}~\bibnamefont {Brandbyge}},\ and\ \bibinfo
  {author} {\bibfnamefont {P.}~\bibnamefont {Hedeg{\aa}rd}},\ }\href
  {https://doi.org/10.1088/0143-0807/35/6/065004} {\bibfield  {journal}
  {\bibinfo  {journal} {European Journal of Physics}\ }\textbf {\bibinfo
  {volume} {35}},\ \bibinfo {pages} {065004} (\bibinfo {year}
  {2014})}\BibitemShut {NoStop}%
\bibitem [{\citenamefont {Král}\ and\ \citenamefont
  {Seideman}(2005)}]{Kral2005}%
  \BibitemOpen
  \bibfield  {author} {\bibinfo {author} {\bibfnamefont {P.}~\bibnamefont
  {Král}}\ and\ \bibinfo {author} {\bibfnamefont {T.}~\bibnamefont
  {Seideman}},\ }\href@noop {} {\bibfield  {journal} {\bibinfo  {journal} {J.
  Chem. Phys.}\ }\textbf {\bibinfo {volume} {123}},\ \bibinfo {pages} {184702}
  (\bibinfo {year} {2005})}\BibitemShut {NoStop}%
\bibitem [{Note1()}]{Note1}%
  \BibitemOpen
  \bibinfo {note} {Projectiles with a very small energy can be, in principle,
  reflected, but this regime is beyond the scope of this work.}\BibitemShut
  {Stop}%
\bibitem [{\citenamefont {Evers}\ \emph {et~al.}(2020)\citenamefont {Evers},
  \citenamefont {Koryt{\'a}r}, \citenamefont {Tewari},\ and\ \citenamefont {van
  Ruitenbeek}}]{Evers2020}%
  \BibitemOpen
  \bibfield  {author} {\bibinfo {author} {\bibfnamefont {F.}~\bibnamefont
  {Evers}}, \bibinfo {author} {\bibfnamefont {R.}~\bibnamefont {Koryt{\'a}r}},
  \bibinfo {author} {\bibfnamefont {S.}~\bibnamefont {Tewari}},\ and\ \bibinfo
  {author} {\bibfnamefont {J.~M.}\ \bibnamefont {van Ruitenbeek}},\ }\href@noop
  {} {\bibfield  {journal} {\bibinfo  {journal} {Rev. Mod. Phys.}\ }\textbf
  {\bibinfo {volume} {92}},\ \bibinfo {pages} {035001} (\bibinfo {year}
  {2020})}\BibitemShut {NoStop}%
\bibitem [{\citenamefont {Evers}\ \emph {et~al.}(2022)\citenamefont {Evers},
  \citenamefont {Aharony}, \citenamefont {Bar-Gill}, \citenamefont
  {Entin-Wohlman}, \citenamefont {Hedeg{\aa}rd}, \citenamefont {Hod},
  \citenamefont {Jelinek}, \citenamefont {Kamieniarz}, \citenamefont
  {Lemeshko}, \citenamefont {Michaeli} \emph {et~al.}}]{Evers2022}%
  \BibitemOpen
  \bibfield  {author} {\bibinfo {author} {\bibfnamefont {F.}~\bibnamefont
  {Evers}}, \bibinfo {author} {\bibfnamefont {A.}~\bibnamefont {Aharony}},
  \bibinfo {author} {\bibfnamefont {N.}~\bibnamefont {Bar-Gill}}, \bibinfo
  {author} {\bibfnamefont {O.}~\bibnamefont {Entin-Wohlman}}, \bibinfo {author}
  {\bibfnamefont {P.}~\bibnamefont {Hedeg{\aa}rd}}, \bibinfo {author}
  {\bibfnamefont {O.}~\bibnamefont {Hod}}, \bibinfo {author} {\bibfnamefont
  {P.}~\bibnamefont {Jelinek}}, \bibinfo {author} {\bibfnamefont
  {G.}~\bibnamefont {Kamieniarz}}, \bibinfo {author} {\bibfnamefont
  {M.}~\bibnamefont {Lemeshko}}, \bibinfo {author} {\bibfnamefont
  {K.}~\bibnamefont {Michaeli}}, \emph {et~al.},\ }\href@noop {} {\bibfield
  {journal} {\bibinfo  {journal} {Advanced Materials}\ }\textbf {\bibinfo
  {volume} {34}},\ \bibinfo {pages} {2106629} (\bibinfo {year}
  {2022})}\BibitemShut {NoStop}%
\bibitem [{\citenamefont {Landau}\ and\ \citenamefont
  {Lifshitz}(1976)}]{Landau}%
  \BibitemOpen
  \bibfield  {author} {\bibinfo {author} {\bibfnamefont {L.~D.}\ \bibnamefont
  {Landau}}\ and\ \bibinfo {author} {\bibfnamefont {E.~M.}\ \bibnamefont
  {Lifshitz}},\ }\href@noop {} {\emph {\bibinfo {title} {Mechanics, Third
  Edition: Volume 1 (Course of\- The\-o\-retical\- Physics)}}},\ \bibinfo
  {edition} {3rd}\ ed.\ (\bibinfo  {publisher} {Butterworth-Heinemann},\
  \bibinfo {year} {1976})\BibitemShut {NoStop}%
\end{thebibliography}%

\end{document}